\documentclass[]{spie}

\usepackage[]{graphicx}

\title{\Large \bf The Fibonacci Model and the Temperley-Lieb Algebra}

\author{Louis H. Kauffman\supit{a} and Samuel J. Lomonaco Jr.\supit{b}
\skiplinehalf
\supit{a} Department of Mathematics, Statistics and Computer Science  
(m/c 249), 851 South Morgan Street, University of Illinois at Chicago,
Chicago, Illinois 60607-7045, USA \\
\supit{b} Department of Computer Science and Electrical Engineering, University of
Maryland Baltimore County, 1000 Hilltop Circle, Baltimore, MD 21250, USA}
 
\authorinfo{Further author information: L.H.K.  E-mail: kauffman@uic.edu,
S.J.L. Jr.: E-mail: lomonaco@umbc.edu}
 
\begin{document} 

  \maketitle

\begin{abstract}
 We give an elementary construction of the Fibonacci model, a unitary braid group representation that is universal for quantum computation.
This paper is dedicated to Professor C. N. Yang, on his $85$-th birthday.
\end{abstract}


\section{Introduction}
This paper gives an elementary construction for the unitary representation of the Artin braid group that constructs the well-known
Fibonacci model. This model gives a topological basis for quantum computation. The present paper is an outgrowth of 
our papers \cite{KLSpie,KLSpie1,Anyonic} and is related to the analysis of the quantum algorithms for the Jones polynomial in the paper by 
Shor and Jordan \cite{Shor}.
We show that unitary representations of the braid group arise naturally in the context of the Temperely - Lieb algebra. 
The Fibonacci model is usually constructed by using recoupling theory. Here we show how the model emerges naturally from braid group representations 
to the Temperley-Lieb algebra. We use the idea of recoupling and change of basis in process spaces to motivate the construction, but we do not have
to rely on any machinery of recoupling theory. This makes the present paper self-contained.
\bigbreak

For the reader
interested in the relevant background in topological quantum computing we recommend the following references 
\{ \cite{Witten,F,FR98,FLZ,F5,F6,Kitaev,MR,Preskill,Wilczek,Ah1,MooreRead} \}.
\bigbreak

Here is a very condensed presentation of how unitary representations of the
braid group are constructed via topological quantum field theoretic methods.
For simplicity assmue that one has a single (mathematical) particle with label $P$
that can interact with itself to produce either itself labeled $P,$ or itself
with the null label $*.$ When $*$ interacts with $P$ the result is always $%
P. $ When $*$ interacts with $*$ the result is always $*.$ One considers
process spaces where a row of particles labeled $P$ can successively
interact, subject to the restriction that the end result is $P.$ For example
the space $V[(ab)c]$ denotes the space of interactions of three particles
labeled $P.$ The particles are placed in the positions $a,b,c.$  Thus we
begin with $(PP)P.$ In a typical sequence of interactions, the first two $P$%
's interact to produce a $*,$ and the $*$ interacts with $P$ to produce $P.$ 
\[
(PP)P \longrightarrow (*)P \longrightarrow P. 
\]
\noindent In another possibility, the first two $P$'s interact to produce a $%
P,$ and the $P$ interacts with $P$ to produce $P.$ 
\[
(PP)P \longrightarrow (P)P \longrightarrow P. 
\]
It follows from this analysis that the space of linear combinations of
processes $V[(ab)c]$ is two dimensional. The two processes we have just
described can be taken to be the the qubit basis for this space. One obtains
a representation of the three strand Artin braid group on $V[(ab)c]$ by
assigning appropriate phase changes to each of the generating processes. One
can think of these phases as corresponding to the interchange of the
particles labeled $a$ and $b$ in the association $(ab)c.$ The other operator
for this representation corresponds to the interchange of $b$ and $c.$ This
interchange is accomplished by a {\it unitary change of basis mapping} 
\[
F:V[(ab)c] \longrightarrow V[a(bc)]. 
\]
\noindent If 
\[
A:V[(ab)c] \longrightarrow V[(ba)c:d] 
\]
is the first braiding operator (corresponding to an interchange of the first
two particles in the association) then the second operator 
\[
B:V[(ab)c] \longrightarrow V[(ac)b] 
\]
is accomplished via the formula $B = F^{-1}AF$ where the $A$ in this formula
acts in the second vector space $V[a(bc)]$ to apply the phases for the
interchange of $b$ and $c.$ \bigbreak

In this scheme, vector spaces corresponding to associated strings of
particle interactions are interrelated by {\it recoupling transformations}
that generalize the mapping $F$ indicated above. A full representation of
the Artin braid group on each space is defined in terms of the local
intechange phase gates and the recoupling transfomations. These gates and
transformations have to satisfy a number of identities in order to produce a
well-defined representation of the braid group. These identities were
discovered originally in relation to topological quantum field theory. In
our approach the structure of phase gates and recoupling
transformations arise naturally from the structure of the bracket model for
the Jones polynomial \cite{JO} and a corresponding representation of the Temperley-Lieb algebra. 
Thus we obtain an entry into  topological
quantum computing that is directly related to the original construction of the Jones polynomial. \bigbreak
\bigbreak

The present paper has two sections. The first section discusses the structure of the Temperley-Lieb algebra in relation 
to the Jones polynomial and the bracket polynomial model for the Jones polynomial. The second section constructs the Fibonacci
model via a representation of the Temperley-Lieb algebra.
\bigbreak

\section{The Bracket Polynomial and the Jones Polynomial}
We now discuss the Jones polynomial. We shall construct the Jones polynomial by using the bracket state 
summation model \cite{KA87}. The bracket polynomial, invariant under Reidmeister moves II and III, can be normalized to give an invariant of all
three Reidemeister moves. This normalized invariant, with a change of variable, is the Jones polynomial
\cite{JO}. The Jones polynomial was originally discovered by a different method than the one given here. 
\bigbreak 

The {\em bracket polynomial} \cite{KA87} , $<K> \, = \, <K>(A)$,  assigns to each unoriented link diagram $K$ a 
Laurent polynomial in the variable $A$, such that
   
\begin{enumerate}
\item If $K$ and $K'$ are regularly isotopic diagrams, then  $<K> \, = \, <K'>$.
  
\item If  $K \sqcup O$  denotes the disjoint union of $K$ with an extra unknotted and unlinked 
component $O$ (also called `loop' or `simple closed curve' or `Jordan curve'), then 

$$< K \sqcup O> \, = \delta<K>,$$ 
where  $$\delta = -A^{2} - A^{-2}.$$
  
\item $<K>$ satisfies the following formulas 

$$<\mbox{\large $\chi$}> \, = A <\mbox{\large $\asymp$}> + A^{-1} <)(>$$
$$<\overline{\mbox{\large $\chi$}}> \, = A^{-1} <\mbox{\large $\asymp$}> + A <)(>,$$
\end{enumerate}

\noindent where the small diagrams represent parts of larger diagrams that are identical except  at
the site indicated in the bracket. We take the convention that the letter chi, \mbox{\large $\chi$},
denotes a crossing where {\em the curved line is crossing over the straight
segment}. The barred letter denotes the switch of this crossing, where {\em the curved
line is undercrossing the straight segment}.  See Figure 1 for a graphic illustration of this relation, and an
indication of the convention for choosing the labels $A$ and $A^{-1}$ at a given crossing.

\begin{figure}
     \begin{center}
     \begin{tabular}{c}
     \includegraphics[height=7cm]{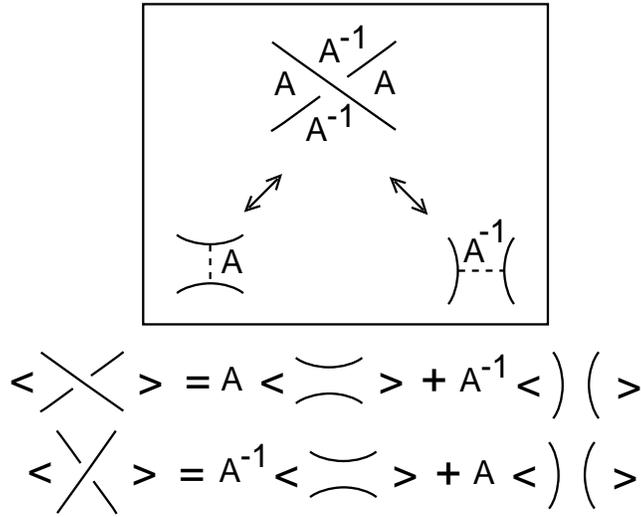}
     \end{tabular}
     \end{center}
     \caption{\bf Bracket Smoothings}
     \end{figure} 
     \bigbreak

\noindent It is easy to see that Properties $2$ and $3$ define the calculation of the bracket on
arbitrary link diagrams. The choices of coefficients ($A$ and $A^{-1}$) and the value of $\delta$
make the bracket invariant under the Reidemeister moves II and III. Thus
Property $1$ is a consequence of the other two properties. 
\bigbreak

In computing the bracket, one finds the following behaviour under Reidemeister move I: 
  $$<\mbox{\large $\gamma$}> = -A^{3}<\smile> \hspace {.5in}$$ and 
  $$<\overline{\mbox{\large $\gamma$}}> = -A^{-3}<\smile> \hspace {.5in}$$

\noindent where \mbox{\large $\gamma$}  denotes a curl of positive type as indicated in Figure 2, 
and  $\overline{\mbox{\large $\gamma$}}$ indicates a curl of negative type, as also seen in this
figure. The type of a curl is the sign of the crossing when we orient it locally. Our convention of
signs is also given in Figure 2. Note that the type of a curl  does not depend on the orientation
we choose.  The small arcs on the right hand side of these formulas indicate
the removal of the curl from the corresponding diagram.  

\bigbreak
  
\noindent The bracket is invariant under regular isotopy and can be  normalized to an invariant of
ambient isotopy by the definition  
$$f_{K}(A) = (-A^{3})^{-w(K)}<K>(A),$$ where we chose an orientation for $K$, and where $w(K)$ is 
the sum of the crossing signs  of the oriented link $K$. $w(K)$ is called the {\em writhe} of $K$. 
The convention for crossing signs is shown in  Figure 2.

\begin{figure}
     \begin{center}
     \begin{tabular}{c}
     \includegraphics[height=3cm]{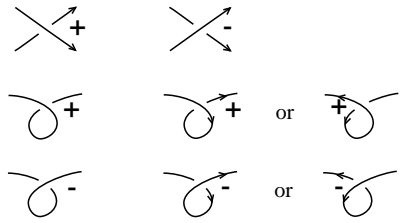}
     \end{tabular}
     \end{center}
     \caption{\bf Crossings Signs and Curls}
     \end{figure} 
     \bigbreak

\noindent {\bf Remark.} By a change of variables one obtains the original
Jones polynomial, $V_{K}(t),$  for oriented knots and links from the normalized bracket:

$$V_{K}(t) = f_{K}(t^{-\frac{1}{4}}).$$

The bracket model for the Jones polynomial is quite useful both theoretically and in terms
 of practical computations. One of the neatest applications is to simply compute $f_{K}(A)$ for the
trefoil knot $K$ and determine that  $f_{K}(A)$ is not equal to $f_{K}(A^{-1}) = f_{-K}(A).$  That computation 
shows that the trefoil is not ambient isotopic to its mirror image, a fact that is much harder to
prove by classical methods.
\bigbreak

\noindent {\bf The State Summation.} In order to obtain a closed formula for the bracket, we now describe it as a state summation.
Let $K$ be any unoriented link diagram. Define a {\em state}, $S$, of $K$  to be a choice of
smoothing for each  crossing of $K.$ There are two choices for smoothing a given  crossing, and
thus there are $2^{N}$ states of a diagram with $N$ crossings.
 In a state we label each smoothing with $A$ or $A^{-1}$ according to the left-right convention 
discussed in Property $3$ (see Figure 1). The label is called a {\em vertex weight} of the state.
There are two evaluations related to a state. The first one is the product of the vertex weights,
denoted  

$$<K|S>.$$
The second evaluation is the number of loops in the state $S$, denoted  $$||S||.$$
  
\noindent Define the {\em state summation}, $<K>$, by the formula 

$$<K> \, = \sum_{S} <K|S>\delta^{||S||-1}.$$
It follows from this definition that $<K>$ satisfies the equations
  
$$<\mbox{\large $\chi$}> \, = A <\mbox{\large $\asymp$}> + A^{-1} <)(>,$$
$$<K \sqcup  O> \, = \delta<K>,$$
$$<O> \, =1.$$
  
\noindent The first equation expresses the fact that the entire set of states of a given diagram is
the union, with respect to a given crossing, of those states with an $A$-type smoothing and those
 with an $A^{-1}$-type smoothing at that crossing. The second and the third equation
are clear from the formula defining the state summation. Hence this state summation produces the
bracket polynomial as we have described it at the beginning of the  section. 

\bigbreak

\noindent{\bf The Temperley Lieb Algebra}
The Temperely Lieb algebra $TL_{n}$ is an algebra over the ring $Z[A, A^{-1}]$ and $\delta = -A^2 - A^{-2}$ with multiplicative generators
$\{ I, U_{1},U_{2},\cdots, U_{n-1} \}$ where $I$ is an identity element and the other generators satisfy the relations
$$U_{i}^{2} =  \delta U_{i},$$
$$U_{i} U_{i \pm 1} U_{i} = U_{i},$$
$$U_{i}U_{j} = U_{j}U_{i},$$ where $|i-j|>1$ in the last equation, and $i$ runs through all values from $1$ to $n-1$ for the first two equations
whenever they are defined. The additive structure of the Temperely-Lieb algebra makes it a free module over the base ring.
\bigbreak

The Temperley-Lieb algebra $TL_{n}$ can be interpreted as shown in Figure 3 in terms of planar connection patterns between two rows of $n$ points.
In this Figure we illustrate the multiplicative relations and we show how a closure of a multiplicative element of the algebra leads to a collection of 
loops in the plane.
There is a trace function $tr:TL_{n} \longrightarrow Z[A, A^{-1}]$ defined in the diagrammatic interpretation by the formula
$$tr(P) = \delta^{||P|| - 1}$$ where $P$ is a product of gererators of the Temperley-Lieb algebra, and $||P||$ denotes the number of loops in the 
closure of the diagrammatic version of $P$, as illustrated in Figure 3. This trace function is extended linearly to the whole algebra, and it has
the property that $tr(ab) = tr(ba)$ for any elements $a$ and $b$ of the algebra.
\bigbreak
 
Another way to think about the bracket polynomial is to first make a representation of the $n$-strand Artin braid group to the 
Temperley-Lieb algebra $TL_{n}$ on $n$-strands and then take the trace ($tr$ as defined above) of this representation.
The representation $$rep: B_{n} \longrightarrow TL_{n}$$ is given by the formula $$rep(\sigma_{i}) = A I + A^{-1} U_{i}$$ where $\sigma_{i}$
denotes the $i$-th braid generator and $I$ denotes the identity element in the Temperley-Lieb algebra. In Figure 3 we have illustrated this formula in
the case
$i=1.$ This illustration should suffice for the  reader to see what is our orientation convention for braid generators, and our convention for the standard
closure of a diagrammatic element of the Temprely-Lieb algebra. This same notion of closure applies to braids. The reader will note that this braid
representation parallels the definition of  the bracket polynomial, so that the expansion formula for the bracket leads directly to the braid
representation when applied to a diagram for the  braid. The following Theorem is a consequence of this correspondence.
\smallbreak

\noindent {\bf Theorem 1.} Let $b$ be a braid in $B_{n}$ and let $\bar{b}$ denote its standard closuure. Let $rep$ denote the representation of the 
braid group discussed above, and let $tr$ denote the trace on the Temperley-Lieb algebra discussed above. Then the bracket polynomial for the braid
closure is given by the formula $$<\bar{b}> = tr(rep(b)).$$
\bigbreak

\begin{figure}
     \begin{center}
     \begin{tabular}{c}
     \includegraphics[height=7cm]{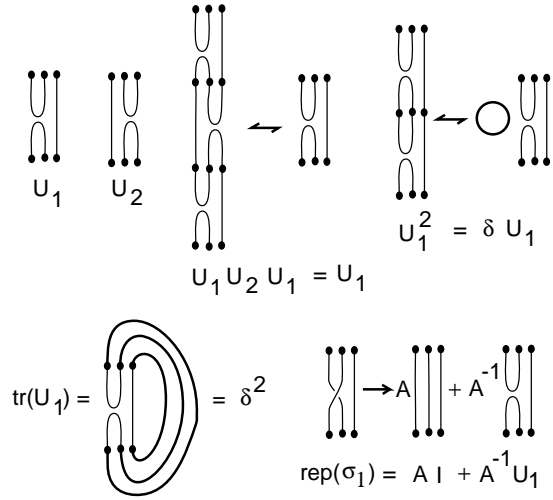}
     \end{tabular}
     \end{center}
     \caption{\bf Diagrammatic Temperley-Lieb Algeba}
     \end{figure} 
     \bigbreak

\noindent {\bf Remark.} We can now explain how to produce unitary representations of the braid group in relation to the Temperley-Lieb algebra and the
Jones polynomial. In order to make a unitary representation of the braid group, it is sufficient to find a representation
$\Gamma: TL_{n} \longrightarrow Aut(V)$ of the Temperley-Lieb algebra to a complex vector space $V$ such that $\Gamma(U_{i})$ is a real and symmetric.
Then, for any element $A = e^{i\theta}$ on the unit circle in the complex plane, we see that $\rho = \Gamma \circ rep$ is a unitary representaton of the
Artin braid group. This statement follows from the fact, that under the above conditions $A I + A^{-1} \Gamma(U_{i})$ is unitary.
In the next section we give an elementary construction for a large class of unitary representations of the three-strand braid group, and then extend this
class to include the Fibonacci model discussed in the Introduction.
\bigbreak

\section {Unitary Representations of the Braid Group and The Fibonacci Model}
The constructions in this section are based on the combinatorics of the Fibonacci model. In this model we have a (mathematical) particle $P$
that interacts with itself either to produce $P$ or to produce a neutral particle $*$. If $X$ is any particle then $*$ iteracts with $X$ to 
produce $X.$ Thus $*$ acts as an identity trasformation. These rules of interaction are illustrated in Figure 4.

\begin{figure}
     \begin{center}
     \begin{tabular}{c}
     \includegraphics[height=4cm]{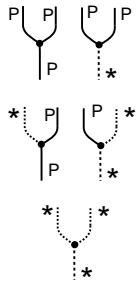}
     \end{tabular}
     \end{center}
     \caption{\bf The Fibonacci Particle $P$}
     \end{figure} 
     \bigbreak

\begin{figure}
     \begin{center}
     \begin{tabular}{c}
     \includegraphics[height=4cm]{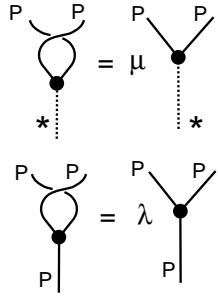}
     \end{tabular}
     \end{center}
     \caption{\bf Local Braiding}
     \end{figure} 
     \bigbreak

The braiding of two particles is measured in relation to their interaction. In Figure 5 we illustrate braiding of $P$ with itself in relation
to the two possible interactions of $P$ with itself. If $P$ interacts to produce $*$, then the braiding gives a phase factor of $\mu.$ If $P$
interacts to produce $P$, then the braiding gives a phase factor of $\lambda.$ We assume at the outset that $\mu$ and $\lambda$ are unit complex
numbers. One should visualize these particles as moving in a plane and the diagrams of interaction are either creations of two particles from one particle,
or fusions of two particles to a single particle (depending on the choice of temporal direction). Thus we have a braiding matrix for these ``local"
particle interactions: 
$$R = 
\left( \begin{array}{cc}
\mu & 0 \\
0 & \lambda \\
\end{array} \right)$$ written with respect to the basis $\{ |* \rangle, |P \rangle \}$ for this space of particle interactions.
\bigbreak

We want to make this braiding matrix part of a larger representation of the braid group. In particular, we want a representation of the three-strand 
braid group on the process space $V_{3}$ illustrated in Figure 6. This space starts with three $P$ particles and considers processes associated
in the patttern $(PP)P$ with the stipulation that the end product is $P$. The possible pathways are illustrated in Figure 6. They correspond
to $(PP)P \longrightarrow (*)P \longrightarrow P$ and $(PP)P \longrightarrow (P)P \longrightarrow P.$ This process space has dimension two and can
support a second braiding generator for the second two strands on the top of the tree. In order to articulate the second braiding we change basis to the
process space corresponding to $P(PP)$ as shown in Figures 7 and 8. The change of basis is shown in Figure 7 and has matrix $F$ as shown below.
We want a unitary representation $\rho$ of three-strand braids so that $\rho(\sigma_{1}) = R$ and $\rho(\sigma_{2}) = S = F^{-1}RF.$ See Figure 8.
We take the form of the matrix $F$ as follows.
$$F = 
\left( \begin{array}{cc}
a & b \\
b & -a \\
\end{array} \right)$$ where $a^2 + b^2 = 1$ with $a$ and $b$ real. This form of the matrix for the basis change is determined by the requirement
that $F$ is symmetric with  $F^{2} = I$. The symmetry of the change of basis formula essentially demands that
$F^{2} = I.$  If $F$ is real, symmetric and $F^2 = I$, then $F$ is unitary.  Since
$R$ is unitary we see that $S = FRF$ is also unitary. Thus, if $F$ is constructed in this way then we obtain a unitary representation of $B_{3}.$
\bigbreak

Now we try to simultaneously construct an $F$ {\it and} construct a representation of the Temperley-Lieb algebra as described in section 2.
We begin by noting that 
$$R = 
\left( \begin{array}{cc}
\mu & 0 \\
0 & \lambda \\
\end{array} \right)
=\left( \begin{array}{cc}
\lambda & 0 \\
0 & \lambda \\
\end{array} \right) + 
\left( \begin{array}{cc}
\mu - \lambda & 0 \\
0 & 0 \\
\end{array} \right) =
\left( \begin{array}{cc}
\lambda & 0 \\
0 & \lambda \\
\end{array} \right) + 
\lambda^{-1} \left( \begin{array}{cc}
\delta & 0 \\
0 & 0 \\
\end{array} \right)$$ where $\delta = \lambda (\mu - \lambda).$
Thus $R = \lambda I + \lambda^{-1} U$ where 
$U = 
\left( \begin{array}{cc}
\delta & 0 \\
0 & 0 \\
\end{array} \right)$ so that $U^2 = \delta U.$ For the Temperley-Lieb representation, we want
$\delta = -\lambda^{2} - \lambda^{-2}$ as explained in section 2. Hence we need
$- \lambda^{2} - \lambda^{-2} = \lambda (\mu - \lambda),$ which implies that $\mu = - \lambda^{-3}.$
With this restriction on $\mu,$ we have the Temperley-Lieb representation and the corresponding unitary braid group representation 
for $2$-strand braids and the $2$-strand Temperley-Lieb algebra.
\bigbreak

\begin{figure}
     \begin{center}
     \begin{tabular}{c}
     \includegraphics[height=2cm]{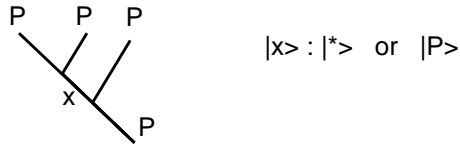}
     \end{tabular}
     \end{center}
     \caption{\bf Three Strands at Dimension Two}
     \end{figure} 
     \bigbreak

\begin{figure}
     \begin{center}
     \begin{tabular}{c}
     \includegraphics[height=4cm]{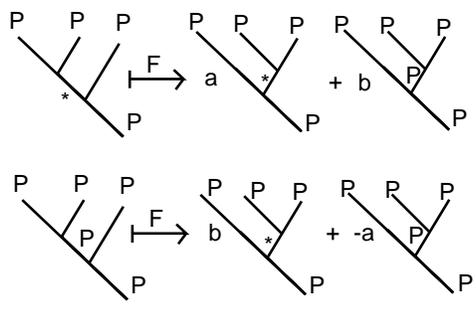}
     \end{tabular}
     \end{center}
     \caption{\bf Recoupling Formula}
     \end{figure} 
     \bigbreak

\begin{figure}
     \begin{center}
     \begin{tabular}{c}
     \includegraphics[height=8cm]{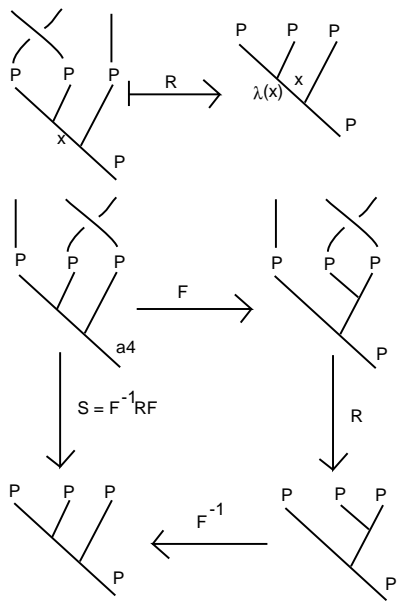}
     \end{tabular}
     \end{center}
     \caption{\bf Change of Basis}
     \end{figure} 
     \bigbreak

Now we can go on to $B_{3}$ and $TL_{3}$ via $S = FRF = \lambda I + \lambda^{-1} V$ with $V = FUF.$ We must examine
$V^{2}$, $UVU$ and $VUV.$ We find that 
$$V^{2} = FUFFUF = FU^{2}F = \delta FUF = \delta V,$$ as desired 	and
$$V = FUF = 
\left( \begin{array}{cc}
a & b \\
b & -a \\
\end{array} \right)
\left( \begin{array}{cc}
\delta & 0 \\
0 & 0 \\
\end{array} \right)
\left( \begin{array}{cc}
a & b \\
b & -a \\
\end{array} \right) =
\delta \left( \begin{array}{cc}
a^2 & ab \\
ab & b^2 \\
\end{array} \right).$$
Thus $V^2 = V$ and since $V = \delta |v\rangle \langle v|$ and $U =\delta |w\rangle \langle w|$ with $w = (1,0)^T$ and $v = Fw = (a,b)^T$
($T$ denotes transpose), we see that 
$$VUV = \delta^3 |v\rangle \langle v|w\rangle \langle w|v\rangle \langle v| = \delta^3 a^2 |v\rangle  \langle v|= \delta^2 a^2 V.$$
Similarly $UVU = \delta^2 a^2 U.$ Thus, we need $\delta^2 a^2 = 1$ and so we shall take $a = \delta^{-1}.$ With this choice, we have 
a representation of the Temperley-Lieb algebra $TL_{3}$ so that $\sigma_{1} = A I + A^{-1} U$ and 
$\sigma_{2} = A I + A^{-1} V$ gives a unitary representation of the braid group when $A = \lambda = e^{i \theta}$ and 
$b = \sqrt{1 - \delta^{-2}}$ is real. This last reality condition is equivalent to the inequality
$$cos^{2}(2 \theta) \ge \frac{1}{4},$$ which is satisfied for infinitely many values of $\theta$ in the ranges 
$$[0,\pi/6] \cup [\pi/3, 2 \pi/3] \cup [5 \pi/6, 7 \pi /6] \cup [4 \pi/3, 5\pi/3].$$
\bigbreak

\noindent With these choices we have 
$$F = 
\left( \begin{array}{cc}
a &  b\\
b & -a \\
\end{array} \right)=
\left( \begin{array}{cc}
1/\delta &  \sqrt{1 - \delta^{-2}}\\
\sqrt{1 - \delta^{-2}} & -1/\delta \\
\end{array} \right)$$ real and unitary, and for the Temperley-Lieb algebra, 
$$U=
\left( \begin{array}{cc}
\delta &  0\\
0 & 0\\
\end{array} \right), 
V = 
\delta \left( \begin{array}{cc}
a^2 &  ab\\
ab & b^2 \\
\end{array} \right) =
\left( \begin{array}{cc}
a &  b\\
b & \delta b^2 \\
\end{array} \right).$$
\bigbreak

Now examine Figure 9. Here we illustrate the action of the braiding and the Temperley-Lieb Algebra on the first Fibonacci 
process space with basis $\{ |*\rangle, |P\rangle \}.$ Here we have $\sigma_{1} = R, \sigma_{2} = FRF$ and $U_{1} =U, U_{2} = V$ as 
described above. Thus we have a representation of the braid group on three strands and a representation of the Temperley-Lieb algebra
on three strands with no further restrictions on $\delta.$
\bigbreak

\begin{figure}
     \begin{center}
     \begin{tabular}{c}
     \includegraphics[height=6cm]{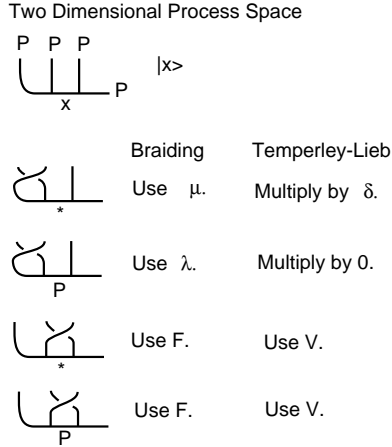}
     \end{tabular}
     \end{center}
     \caption{\bf Algebra for a Two Dimensional Process Space}
     \end{figure} 
     \bigbreak

\begin{figure}
     \begin{center}
     \begin{tabular}{c}
     \includegraphics[height=3cm]{F10.epsf}
     \end{tabular}
     \end{center}
     \caption{\bf A Five Dimensional Process Space}
     \end{figure} 
     \bigbreak

\begin{figure}
     \begin{center}
     \begin{tabular}{c}
     \includegraphics[height=7cm]{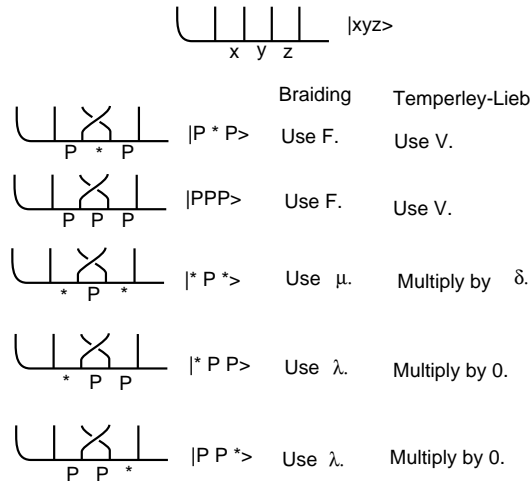}
     \end{tabular}
     \end{center}
     \caption{\bf  Algebra for a Five Dimensional Process Space}
     \end{figure} 
     \bigbreak

So far, we have arrived at exactly the $3$-strand braid representations that we used in our papers \cite{QCJP1,QCJP2} giving a quantum algorithm for the
Jones polynomial for three-strand braids. In this paper we are working in the context of the Fibonacci process spaces and so we wish to see how to make a 
representation of the Temperley-Lieb algebra to this model as a whole, not restricting ourselves to only three strands. The generic case to
consider is the action of the Temperley-Lieb algebra on process spaces of higher dimension as shown in Figures 10 and 11. In the Figure 11 we
have illustrated the triplets from the previous figure as part of a possibly larger tree and have drawn the strings horizontally rather than diagonally. In
this figure we have listed the effects of braiding the vertical strands $3$ and $4$. We see from this figure that the action of the Temperley-Lieb algebra
must be as follows:
$$U_{3}|P*P\rangle = a|P*P\rangle + b|PPP\rangle,$$
$$U_{3}|PPP\rangle = b|P*P\rangle + \delta b^{2} |PPP\rangle,$$
$$U_{3}|*P*\rangle = \delta |*P*\rangle,$$
$$U_{3}|*PP\rangle = 0,$$
$$U_{3}|PP*\rangle = 0.$$
Here we have denoted this action as $U_{3}$ because it connotes the action on the third and fourth vertical strands in the sequences shown in Figure 11.
Note that in a larger sequence we can recognize $U_{j}$ by examining the triplet surrounding the $j-1$-th element in the sequence, just as the pattern 
above is governed by the elements surrounding the second element in the sequence. For simplicity, we have only indicated three elements in the sequences
above. Note that in a sequence for the Fibonacci process there are never two consecutive appearances of the neutral element $*.$ 
\bigbreak

We shall refer to a sequence of $*$ and $P$ as a {\it Fibonacci sequence} if it contains no consecutive appearances of $*$. Thus
$|PP*P*P*P\rangle$ is a Fibonacci sequence. In working with this representation of the braid group and
Temperley-Lieb algebra, it is convenient to assume that the ends of the sequence are flanked by $P$ as in Figures 10 and 11 for sequences of length
$3.$ It is convenient to leave out the flanking $P$'s when notating the sequence.
\bigbreak

Using these formulas we can determine conditions on $\delta$ such that this is a representation of the Temperley-Lieb algebra for all Fibonacci sequences.
Consider the following calculation:
$$U_{4}U_{3}U_{4}|PPPP\rangle = U_{3}U_{2}(b|PP*P\rangle + \delta b^{2}|PPPP\rangle)$$
$$=  U_{4}(bU_{3}|PP*P\rangle + \delta b^{2}U_{3}|PPPP\rangle)$$
$$=  U_{4}(0 + \delta b^{2}(b|P*PP\rangle + \delta b^{2}|PPPP\rangle)$$
$$=  \delta b^{2}(bU_{4}|P*PP\rangle + \delta b^{2}U_{4}|PPPP\rangle)$$
$$=  \delta^{2} b^{4}U_{4}|PPPP\rangle.$$
Thus we see that in order for $U_{4}U_{3}U_{4} = U_{4},$ we need that $\delta^{2} b^{4} = 1.$ 
\bigbreak

{\it It is easy to see that $\delta^{2} b^{4} = 1$ is the only remaining
condition needed to make sure that the action of the Temperley-Lieb algebra extends to all Fibonacci Model sequences.} 
\bigbreak

\noindent Note that 
$\delta^{2} b^{4} = \delta^{2}(1 - \delta^{-2})^2 = (\delta - 1/\delta)^{2}.$ Thus we require that 
$$\delta - 1/\delta = \pm 1.$$
When $\delta - 1/\delta = 1,$ we have the solutions $\delta = \frac{1 \pm \sqrt{5}}{2}.$ However, for the reality of $F$ we require that
$1 - \delta^{-2} \ge 0,$ ruling out the choice $\delta = \frac{1 - \sqrt{5}}{2}.$ When $\delta - 1/\delta = -1,$ we have
the solutions $\delta = \frac{-1 \pm \sqrt{5}}{2}.$ This leaves only $\delta = \pm \phi$ where $\phi = \frac{1 + \sqrt{5}}{2}$ (the Golden
Ratio) as possible values for $\delta$ that satisfy the reality condition for $F.$ Thus, up to a sign we have arrived at the well-known value of
$\delta = \phi$ (the Fibonacci model) as essentially the only way to have an extension of this form of the representation of the Temperley-Lieb
algebra for $n$ strands. Let's state this positively as a Theorem.
\bigbreak

\noindent {\bf Theorem 2.} Let $V_{n+2}$ be the complex vector space with basis $\{ |x_{1}x_{2}\cdots x_{n} \rangle \}$ where each $x_{i}$ equals
either $P$ or $*$ and there do {\it not} occur two consecutive appearances  of $*$ in the sequence $\{ x_{1},\cdots x_{n} \}.$ We refer to this basis for
$V_{n}$ as the set of {\it Fibonacci sequences} of length $n.$ Then the dimension of $V_{n}$ is equal to $f_{n+1}$ where
$f_{n}$ is the $n$-th Fibonacci number: $f_{0} = f_{1} = 1$ and $f_{n+1} = f_{n} + f_{n-1}.$ Let $\delta = \pm \phi$
where $\phi = \frac{1 + \sqrt{5}}{2}.$ Let $a = 1/\delta$ and $b = \sqrt{1 - a^2}.$ Then the Temperley-Lieb algebra on $n+2$ strands with loop value
$\delta$ acts on $V_{n}$ via the formulas given below. 
First we give the left-end actions.
$$U_{1}|*x_{2}x_{3} \cdots x_{n} \rangle = \delta |*x_{2}x_{3} \cdots x_{n} \rangle,$$
$$U_{1}|Px_{2}x_{3} \cdots x_{n} \rangle = 0,$$
$$U_{2}|*Px_{3} \cdots x_{n} \rangle = a|*P x_{3} \cdots x_{n} \rangle + b|PP x_{3} \cdots x_{n} \rangle, $$
$$U_{2}|P* x_{3} \cdots x_{n} \rangle = 0,$$
$$U_{2}|PP x_{3} \cdots x_{n} \rangle = b|*P x_{3} \cdots x_{n} \rangle + \delta b^{2}|PP x_{3} \cdots x_{n} \rangle.$$
Then we give the general action for the middle strands.
$$U_{i}|x_{1} \cdots x_{i-3} P*P x_{i+1} \cdots x_{n} \rangle = a|x_{1} \cdots x_{i-3} P*P  x_{i+1} \cdots x_{n} \rangle$$
$$+ b|x_{1} \cdots x_{i-3} PPP x_{i+1} \cdots x_{n} \rangle,$$
$$U_{i}|x_{1} \cdots x_{i-3} PPP x_{i+1} \cdots x_{n} \rangle = b|x_{1} \cdots x_{i-3} P*P x_{i+1} \cdots x_{n} \rangle$$ 
$$+ \delta b^{2}|x_{1} \cdots x_{i-3} PPP x_{i+1} \cdots x_{n} \rangle,$$
$$U_{i}|x_{1} \cdots x_{i-3} *P* x_{i+1} \cdots x_{n} \rangle = \delta |x_{1} \cdots x_{i-3} *P* x_{i+1} \cdots x_{n} \rangle,$$
$$U_{i}|x_{1} \cdots x_{i-3} *PP x_{i+1} \cdots x_{n} \rangle = 0,$$
$$U_{i}|x_{1} \cdots x_{i-3} PP* x_{i+1} \cdots x_{n} \rangle = 0.$$
Finally, we give the right-end action.
$$U_{n+1}|x_{1} \cdots x_{n-2} *P \rangle = 0,$$
$$U_{n+1}|x_{1} \cdots x_{n-2} P* \rangle =0,$$
$$U_{n+1}|x_{1} \cdots x_{n-2} PP \rangle =b|x_{1} \cdots x_{n-2} P* \rangle + \delta b^{2}|x_{1} \cdots x_{n-2} PP \rangle.$$
\bigbreak

\noindent {\it Remark.} Note that the left and right end Temperley-Lieb actions depend on the same basic pattern as the middle action.
The Fibonacci sequences $|x_{1} x_{2} \cdots x_{n} \rangle$ should be regarded as flanked left and right by $P$'s just as in the special
cases discussed prior to the proof of Theorem $2$.
\bigbreak

\noindent {\bf Corollary.} With the hypotheses of Theorem $2$, we have a unitary representation of the Artin Braid group $B_{n+2}$ to
$TL_{n+2}$, $\rho: B_{n+2} \longrightarrow TL_{n+2}$ given by the formulas
$$\rho(\sigma_{i}) = A I + A^{-1} U_{i},$$
$$\rho(\sigma_{i}^{-1}) = A^{-1} I + A U_{i},$$ where $A = e^{3 \pi i/5}$
where the $U_{i}$ connote the representation of the Temperley-Lieb algebra on the space $V_{n+2}$ of Fibonacci sequences as described in the Theorem 
above.
\bigbreak

\noindent {\bf Remark.} The Theorem and Corollary give the original parameters of 
the Fibonacci model and shows that this model admits a unitary representation of the braid group via a Jones representation of the Temperley-Lieb
algebra.
\bigbreak

In the original Fibonacci model \cite{Anyonic}, there is a basic non-trivial recoupling matrix $F.$ 
$$F =
\left( \begin{array}{cc}
1/\delta & 1/\sqrt{\delta} \\
1/\sqrt{\delta} & -1/\delta \\
\end{array} \right) =
\left( \begin{array}{cc}
\tau & \sqrt{\tau} \\
\sqrt{\tau} & -\tau \\
\end{array} \right)$$
where $\delta = \frac{1 + \sqrt{5}}{2}$ is the golden ratio and $\tau = 1/\delta$.
The local braiding matrix is given by the formula
below with $A = e^{3\pi i/5}.$
$$R = 
\left( \begin{array}{cc}
A^{8} & 0 \\
0 & -A^{4} \\
\end{array} \right)=
\left( \begin{array}{cc}
e^{4\pi i/5} & 0 \\
0 & -e^{2\pi i/5} \\
\end{array} \right).$$
\bigbreak

This is exactly what we get from our method by using $\delta = \frac{1 + \sqrt{5}}{2}$ and $A = e^{3 \pi i/5}.$
Just as we have explained earlier in this paper, the simplest example of a braid group representation arising from this theory is the representation of the
three strand braid group generated by
$\sigma_{1}= R$ and $\sigma_{2} = FRF$ (Remember that $F=F^{T} = F^{-1}.$). The matrices $\sigma_{1}$ and $\sigma_{2}$ are both unitary, and they generate
a dense subset of $U(2),$ supplying the local unitary transformations needed for quantum computing. The full braid group representation on the Fibonacci
sequences is  computationally universal for quantum computation. In our earlier paper \cite{Anyonic} we gave a construction for the Fibonacci model
based on Temperely-Lieb recoupling theory. In this paper, we have reconstructed the Fibonacci model on the more elementary grounds of the representation
of the Temperley-Lieb algebra summarized in the statement of the Theorem $2$ and its Corollary.
\bigbreak


\begin{thebibliography}{1}
\bibitem{Ah1}
D. Aharonov, V. Jones, Z. Landau,
A polynomial quantum algorithm for approximating the Jones polynomial,
quant-ph/0511096.

\bibitem{F} 
M. Freedman, A magnetic model with a possible Chern-Simons phase,
quant-ph/0110060v1 9 Oct 2001, (2001), preprint

\bibitem{FR98} 
M. Freedman,  {\em Topological Views on Computational Complexity}, Documenta
Mathematica - Extra Volume ICM, 1998, pp. 453--464.

\bibitem{FLZ} 
M. Freedman, M. Larsen, and Z. Wang, A modular functor which is universal for 
quantum computation,  quant-ph/0001108v2, 1 Feb 2000.

\bibitem {F5}
M. H. Freedman,  A. Kitaev, Z. Wang,
Simulation of topological field theories by quantum computers, {\em Commun. Math. Phys.}, {\bf 227}, 587-603 (2002),
quant-ph/0001071.

\bibitem {F6}
M. Freedman,  {Quantum computation and the
localization of modular functors}, quant-ph/0003128.

\bibitem{JO} 
V.F.R. Jones, A polynomial invariant for links via von Neumann algebras,
Bull. Amer. Math. Soc. {\bf 129} (1985), 103--112.

\bibitem{KA87}  
L.H. Kauffman, State models and the Jones polynomial, Topology {\bf 26} (1987),
395--407.

\bibitem{KLSpie} L. H. Kauffman and S. J. Lomonaco Jr., Spin Networks and anyonic topological computing, In ``Quantum Information and Quantum
Computation IV", (Proceedings of Spie, April 17-19,2006) edited by E.J. Donkor, A.R. Pirich and H.E. Brandt,
Volume 6244, Intl Soc. Opt. Eng., pp. 62440Y-1 to 62440Y-12.

\bibitem{KLSpie1} L. H. Kauffman and S. J. Lomonaco Jr., Spin Networks and anyonic topological computing II, In
``Quantum Information and Quantum Computation V", (Proceedings of Spie, April 10-12,2007) edited by E.J.
Donkor, A.R. Pirich and H.E. Brandt, Volume 6573, Intl Soc. Opt. Eng., pp. 65730U-1 to 65730u-13.

\bibitem{Anyonic} L. H. Kauffman and S. J. Lomonaco Jr., $q$ - Deformed Spin Networks, Knot Polynomials and Anyonic Topological Quantum Computation,
 {\it JKTR} Vol. 16, No. 3 (March 2007), pp. 267-332.

\bibitem{QCJP1}
L. H. Kauffman, Quantum computing and the Jones polynomial, math.QA/0105255, in {\em Quantum Computation and Information}, S. Lomonaco, 
Jr. (ed.), AMS CONM/305, 2002, pp.~101--137.

\bibitem{QCJP2}
L.H. Kauffman and Samuel J. Lomonaco Jr. A Three-stranded quantum algorithm for the Jones polynonmial, in
``Quantum Information and Quantum Computation V", (Proceedings of Spie, April 2007) edited by E.J. Donkor, A.R. Pirich and H.E. Brandt,
Intl Soc. Opt. Eng. , 65730T-1-16.

\bibitem{Kitaev}
A. Kitaev, Anyons in an exactly solved model and beyond, 
{\em arXiv.cond-mat/0506438 v1 17 June 2005}.

\bibitem{MooreRead}
G. Moore and N. Read, Non-abelions in the fractional quantum Hall effect, {\it Nuclear PhysicsB} 360 (1991), 362-396.

\bibitem{MR}
A. Marzuoli and M. Rasetti, Spin network quantum simulator, {\em Physics Letters A} {\bf 306} (2002) 79--87.

\bibitem{Preskill}
J. Preskill, Topological computing for beginners, (slide presentation), Lecture Notes for Chapter 9 - Physics 219 - Quantum Computation.
{\it http://www.iqi.caltech.edu/~preskill/ph219}

\bibitem{Shor}
P. W. Shor and S. P. Jordan, Estimating Jones polynomials is a complete problem for one clean qubit.
{\it arxiv:0707.2831v1 [quqnt-ph]} 19 Jul 2007.

\bibitem{Wilczek}
F. Wilczek, {\em Fractional Statistics and Anyon Superconductivity,} World Scientific Publishing Company (1990).

\bibitem{Witten} E. Witten, Quantum field Theory and the Jones Polynomial, {\em
Commun. Math. Phys.},vol. 121, 1989, pp. 351-399.

\end{thebibliography}
 \end{document}